# Interpretable Ensemble Learning for Materials Property Prediction with Classical Interatomic Potentials: Carbon as an Example

Xinyu Jiang[1], Haofan Sun[1], Kamal Choudhary[2], Houlong Zhuang[1,†], and Qiong Nian[1,*]

[1]School for Engineering of Matter, Transport and Energy, Arizona State University, Tempe, AZ 85287, USA
[2]Material Measurement Laboratory, National Institute of Standards and Technology, Gaithersburg, MD, 20899, USA
†zhuanghl@asu.edu
*Qiong.Nian@asu.edu

**Abstract**

Machine learning (ML) is widely used to explore crystal materials and predict their properties. However, the training is time-consuming for deep-learning models, and the regression process is a black box that is hard to interpret. Also, the preprocess to transfer a crystal structure into the input of ML, called descriptor, needs to be designed carefully. To efficiently predict important properties of materials, we propose an approach based on ensemble learning consisting of regression trees to predict formation energy and elastic constants based on small-size datasets of carbon allotropes as an example. Without using any descriptor, the inputs are the properties calculated by molecular dynamics with 9 different classical interatomic potentials. Overall, the results from ensemble learning are more accurate than those from classical interatomic potentials, and ensemble learning can capture the relatively accurate properties from the 9 classical potentials as criteria for predicting the final properties.

# INTRODUCTION

In recent years, density functional theory (DFT) and molecular dynamics (MD) simulations have been studied and applied extensively in materials multiscale modeling [1]. For example, the calculation of energy and forces of materials across different scales have been achieved by using these simulations [2-4]. Current widely used simulation methods including Kohn-Sham density functional theory (KSDFT) [5, 6] and MD simulations with classical interatomic potentials [7-16] have demonstrated high performance in predicting the formation energy and elastic modulus of materials. However, both methods have their own limitations. KSDFT is computationally demanding and typically restricted to systems containing only a few hundred of atoms, while MD can be used in larger systems but is limited in accuracy due to the empirical nature of interatomic potentials.

To solve the limitations of KSDFT and MD, machine learning (ML) models [17, 18] such as neural network potential (NNP) [19, 20], Gaussian approximation potential (GAP) [21], spectral neighbor analysis potential (SNAP) [22, 23], and moment tensor potential (MTP) [24] have been proposed to accurately predict energy and forces of crystals and molecules. They use atomic species and nuclear coordinates to build descriptors (also called "fingerprints"), which are invariant under permutations among the same elements, and isometric transformations of rotations, as features to be fitted by a chosen regression model [19, 25]. However, these descriptors need to be designed meticulously to satisfy the restrictions and the complex transformations thus making it difficult to explain the models [26, 27].

To get more generalized descriptors, graph networks, which represent atoms and bonds as nodes and edges, respectively, combined with convolutional neural networks have received significant attention, since convolutional neural networks can automatically find the important features compared to descriptor-based models [28]. Several graph convolutional neural networks such as generalized crystal graph convolutional neural networks (CGCNN) [26], SchNet [29],MEGNet [30] and ALIGNN [31] have been proposed. They are straightforward to be adopted and suitable for both crystals and molecules, however, these descriptors have complex configurations that contain a series of operators and hidden layers, and their fitting process is time-consuming due to the high data requirements and the regression function contains a large number of parameters to be fitted in the neural network [32].

Compared to graph-network-based potentials, symbolic regression is a faster method to build interatomic potentials by using genetic programming to find a function that accurately expresses interatomic potentials from a set of variables and mathematical operators [33-36]. But symbolic regression also has some limitations. For example, the expressions in the hypothesis space must be simple and have a significant effect on potential energy, and this model cannot learn complex terms that involve bond angles.

Besides, for general ML potentials, transferability, which describes the ability of a model to correctly predict the property of an atomic configuration lying outside the training dataset, is limited. Consequently, physically informed neural networks (PINN) are proposed to improve the transferability of unknown structures [37-39] by combining a general physics-based interatomic potential with a neural-network regression. PINN achieve this by optimizing a set of physical-meaning parameters of a physics-based interatomic potential from trained neural networks, and then feeding them back to improve the accuracy of the original physics-based interatomic potential. However, this method encounters a similar obstacle to the graph networks mentioned above, which is the time-consuming fitting process in the neural network results from a large size of data and numerous parameters.

In this study, we present a regression-trees-based ensemble learning approach that efficiently predicts the formation energy and elastic constants of carbon allotropes with a small size of data. We use carbon allotropes as an example to evaluate the performance of our model since carbon is one of the fundamental elements on Earth [40] these carbon allotropes have a variety of physical properties while being applied widely in cutting and polishing tools [41], superlubricity [42], solar thermal energy storage [43], etc. Therefore, understanding of the physical properties of carbon allotropes plays a significant role in both scientific research and engineering applications. We begin by extracting the structures of carbon allotropes from the Materials Project (MP) [44], and compute their formation energy and elastic constants using MD simulations with 9 different classical interatomic potentials via the Large-scale Atomic/Molecular Massively Parallel Simulator (LAMMPS) [45]. Then use these computed properties as features and corresponding DFT references as targets to train and test four different ensemble learning models [46-49] consisting of regression trees [50]. In general, the performance of ensemble learning models is better than that of 9 classical potentials, and based on feature importance, ensemble learning can find the accurate features and use them to improve the precision of prediction.

RESULTS

**Ensemble learning framework for properties prediction of carbon materials**

Fig. 1 illustrates the schematic of the ensemble learning framework. Firstly, carbon structures are extracted from the MP database. Then, the formation energy and elastic constants of each structure are calculated by MD with 9 classical interatomic potentials, including the analytic bond-order potential (ABOP) [51], adaptive intermolecular reactive empirical bond order potential (AIREBO) [9], standard Lennard-Jones potential (LJ) [13], AIREBO-M [10] potential that replaces the LJ term with a Morse potential in AIREBO potential [52], environment-dependent interaction potential (EDIP) [53], long-range carbon bond order potential (LCBOP) [12], modified embedded atom method (MEAM) [14], reactive force field potential (ReaxFF) [15], and Tersoff potential [54]. The training dataset is composed of these properties and corresponding DFT reference collected from the MP database, and encoded into feature vectors $x_i$ and target vectors $y_i$, respectively. For the formation energy, 58 carbon structures and their DFT reference are extracted from the MP database, and for the elastic constants, 20 out of 58 carbon structures' DFT reference are used due to the absence of DFT reference and removal of unstable or erroneous calculations [55]. Next, the regression-trees-based ensemble models will be trained based on these vectors. We select regression-trees-based ensemble learning models for the following reasons. First, compared to neural networks, regression trees are white-box models which make the models and outputs easy to understand and interpret. Second, as non-linear models, regression trees have better performance than classical linear regression and neural networks methods when dealing with small-size data and highly non-linear features. Third, to mitigate locally optimal decisions of regression trees, ensemble learning combined with the predictions of several regression trees will be deployed to improv robustness over a single regression tree. Last, for multi-target problems such as the prediction of elastic constants, the ensemble learning method can learn the correlations of elastic constants and output multiple targets at once. Here, different regression-trees-based ensemble-learning methods implemented in the Scikit-Learn package [56], including bootstrap aggregation (bagging) [57] and boosting [47], are used to build simple, fast, and interpretable models. Details of the architectures and methods of regression trees and ensemble learning are given in the "Methods" section. For the new carbon structures, we can calculate the same properties by using MD with the 9 potentials and feed these calculated properties into the trained model with the smallest mean absolute error (MAE) during testing, as shown in Eq. 1, to predict the properties of new structures.

$$MAE = \frac{\sum_{i=1}^{n}|y_i^{pre}-y_i|}{n} \tag{1}$$

Where $y_i^{pre}$ is prediction of model, $y_i$ is value of reference, and $n$ is the number of samples.

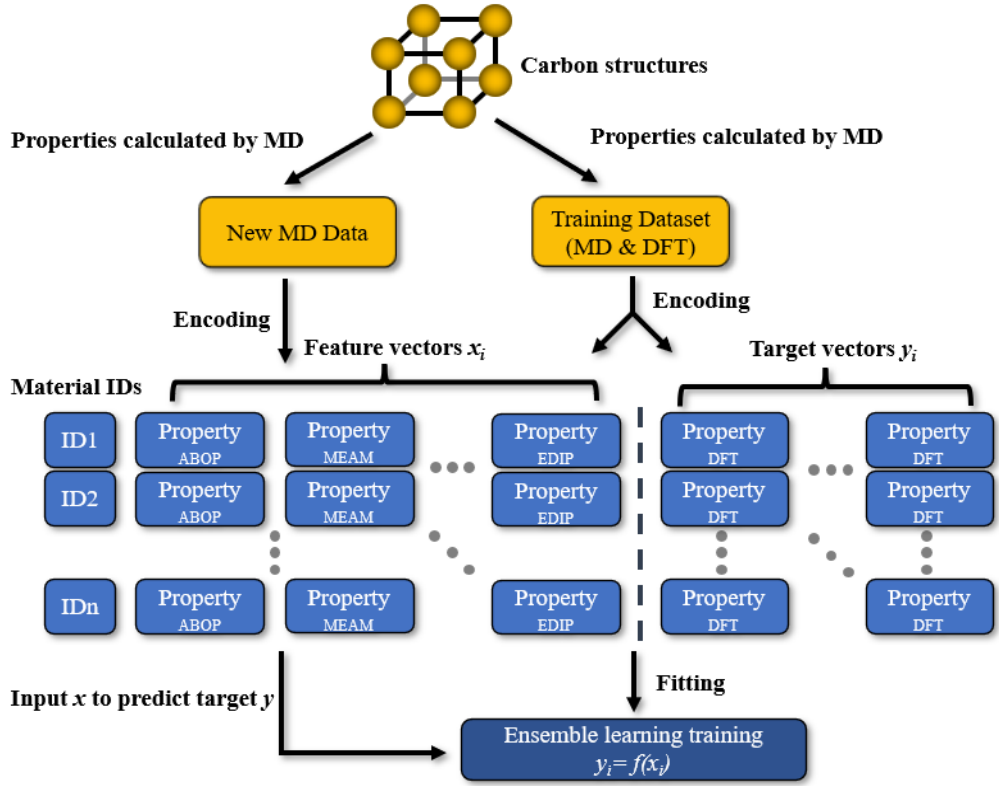

Fig. 1. Ensemble learning framework for properties prediction of carbon materials. The properties of carbon structures are calculated by MD and use these calculated values and DFT references as input to train the ensemble learning model. To get the same properties of new carbon structures, first calculate the properties of new structures by using MD, then put these values into the best trained model to get final predictions.

For the formation energy of carbon materials, we employ four different ensemble-learning methods, namely RandomForest (RF), AdaBoost (AB), GradientBoosting (GB), and XGBoost (XGB), to evaluate their performance. Grid search in combination with 10-fold cross-validation is applied to optimize hyperparameters. After tuning, we run 10-fold cross-validation twenty times for each method with the optimized hyperparameters and calculate the MAEs compared to DFT reference. Furthermore, the median absolute deviation (MAD) of each method is calculated. MAD is defined as the median of the absolute deviations from the median of the residuals, as shown in Eq. 2 and Eq. 3.

$$MAD = median(|r_i - \tilde{r}|) \tag{2}$$

Where $r_i$ is residual between *ith* target's prediction and its' corresponding target. $\tilde{r}$ represents residuals' median.

$$\tilde{r} = median(r) \tag{3}$$

MAD represents the dispersion of residuals. It is more robust than MAE since it can ignore the influence of outliers. The MAE and MAD for each method are depicted in Fig. 2. Here, the voting regressor (VR) which combined RF, AB, and GB models is utilized to mitigate the overall error by averaging the predictions. Besides, Gaussian process (GP) [58], as a generic supervised learning method to solve regression problems, is also evaluated. Overall, ensemble-learning models have better performance than that of classical interatomic potentials and GP model. Notably, all these MAEs are lower than the most accurate classical potential, LCBOP. Since the formation energy values calculated by different classical potentials have high non-linear and complex relationships, under this condition, the regression trees have better performance than classical regression methods such as GP. In the inset of Fig. 2, the formation energy of various structures predicted by RF, LCBOP, and DFT are illustrated. It can be observed that RF outperforms LCBOP in terms of overall error. However, for the structures with the highest formation energy, RF's predictions are less accurate than those by LCBOP may be due to the inherent nature of weak extrapolation that causes the lower formation energy of mp-998866. It is worth noting that RF has weak performance for mp-1008395,

mp-570002, mp-624889, and mp-1018088. The reason is the deviation of the features with respect to DFT for each structure. For mp-1018088, except for the feature calculated by LJ potential, the features calculated by the other potentials are smaller than the reference, leading to an underestimated prediction. The values of mp-624889 have a distribution similar to mp-1018088, but the deviation is smaller (-0.33 eV/atom for mp-624889 and -0.55 eV/atom for mp-1018088) that makes the prediction more accurate. Conversely, since the values of mp-1008395 and mp-570002 calculated by the most classical potentials are larger than the reference, the predictions of both are overestimated. Nevertheless, RF provide more accurate predictions in general.

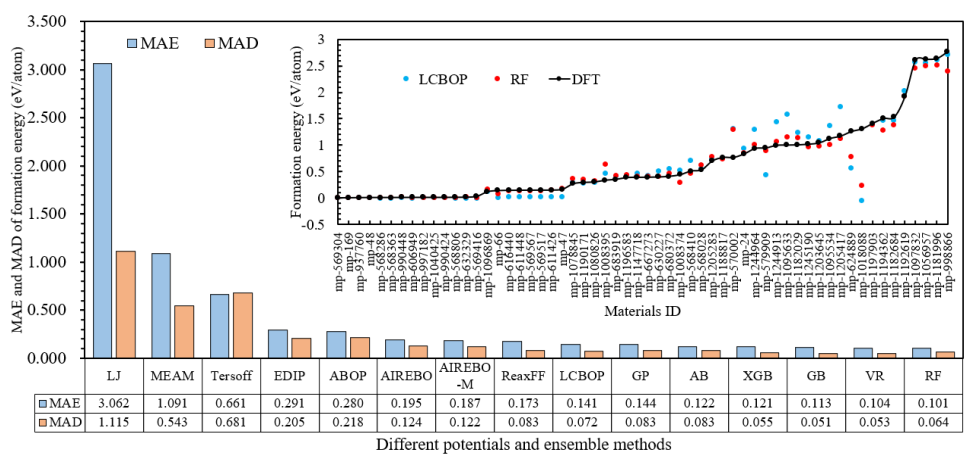

Fig. 2. MAEs and MADs of the formation energy relative to DFT reference under different methods. Overall, ensemble-learning models have better performance than that of classical interatomic potentials and GP model, and the RF has the best performance than the others. Here, the inset shows the prediction of the formation energy of carbon structures by LCBOP and RF. The gray dotted line represents the DFT reference. The RF has better performance than LCBOP overall. For the structures with high formation energy, due to the inherent nature of weak extrapolation, RF is less accurate than LCBOP.

For the elastic constants of carbon materials, we also train and test RF, AB, GB, and XGB models using elastic constants calculated by the same 9 classical potentials. Grid search in combination with 5-fold cross-validation is applied to tune the hyperparameters of each model. Then the 10-fold cross-validation is conducted on the models with optimized hyperparameters to evaluate the performance of the models. The prediction of elastic constants is a multi-target problem. However, AB, GB, and XGB don't support multi-target regression. To overcome this limitation, we use a multi-target regressor [56] combined with all four ensemble methods to predict elastic constants. In brief, the multi-target regressor fits one regressor based on ensemble methods per elastic constant, it is a simple strategy to extend the regressors which don't support multi-target problems. Here, we use the Tersoff potential as a benchmark to compare with different ensemble methods since the MAE of the elastic constants of the Tersoff potential is at least an order of magnitude smaller than those of other classical potentials. Fig. 3a illustrates the MAEs of the total elastic constants of four ensemble methods conducted by twenty times 10-fold cross-validation individually. The MAEs of AB, RF, XGB, and GB are much smaller than that of Tersoff. It is worth noting that the MAE of Tersoff is significantly increased by one structure (mp-1095534) that includes both sp and $sp^3$ hybridization of carbon, making the structure more complex than others such as diamond or graphite, which only have single hybridization. This complexity makes it difficult for Tersoff to accurately calculate. If the error associated with the mp-1095534 is removed, the MAE of Tersoff drops to 63 GPa, which is still larger than those of RF, AB, GB, and XGB. Notably, different potentials behave differently for different carbon structures, hence, to get a minimal MAE of all structures from classical potentials, we extract the smallest error among 9 classical potentials with respect to DFT reference for each structure, and then calculate the total MAE of these smallest errors. As shown in Fig. 3a, the Min represents the best performance by using these classical potentials. We can see that AB has a smaller MAE than Min, and XGB performs similarly to Min. Fig. 3b shows the elastic constants calculated by Tersoff and predicted by AB and RF by using 10-fold cross-validation. The black dashed lines are the ideal fit (1:1). To fit the plot, fifteen points of Tersoff with excessively large MAEs are removed. Both AB and RF have lower MAEs than Tersoff, as shown in Fig. 3a. The AB has a lower MAE than RF, possibly due to the elastic constants of similar structures correlated with each other, and a sequential process like AB can reduce the bias. The points in the green circles in Fig. 3b have large errors, and all these points come from $C_{11}$, $C_{22}$ or $C_{33}$ of the structures that have complicated structures not trained in the models. These structures have smaller values in $C_{11}$, $C_{22}$ or $C_{33}$ compared to most of the training data. Therefore, the models don't have enough fitted regressors for these structures, resulting in inaccurate predictions. To further assess the performance of ensemble methods compared to Tersoff, the MAEs and MADs of partial elastic constants for AB and Tersoff are calculated in Table 1. We exclude the MAEs and MADs of the remaining elastic constants due to their negligible errors. In general, all 9 smallest MAEs and MADs are obtained from AB. For Tersoff with mp-1095534, the MAEs and MADs are higher than the others. Although Tersoff without mp-1095534 yields smaller MAEs and MADs than that with mp-1095534, the MAEs and MADs of AB are still smaller, some are even over 50% lower than those of Tersoff without mp-1095534. Both tables demonstrate that the ensemble learning has better performance than Tersoff.

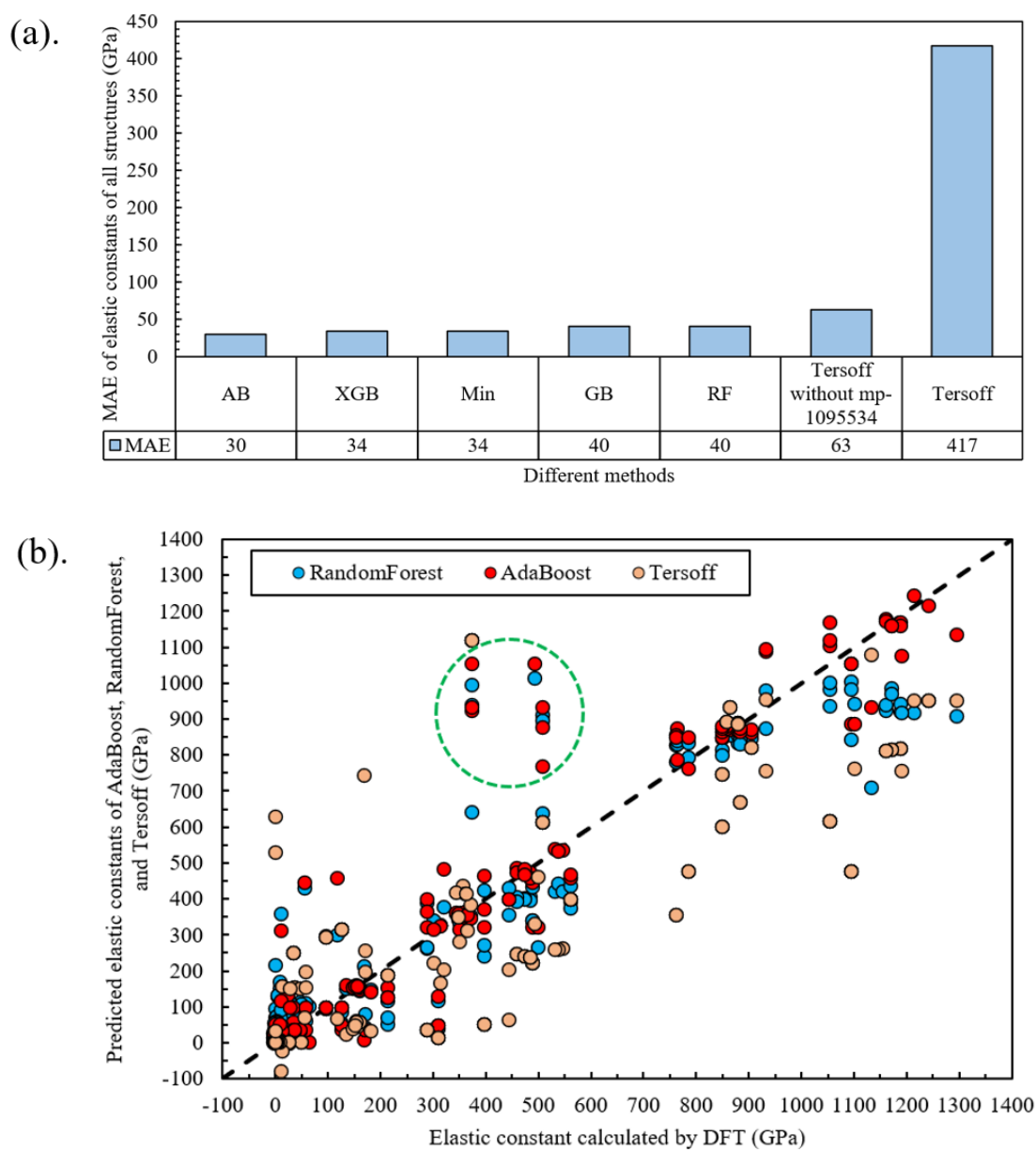

Fig. 3. Performance of different classical potentials and ML methods in elastic constants prediction. (a). MAEs of the total elastic constants relative to DFT reference under different methods, Min represents the best performance for all structures by using these 9 classical potentials. AB and XGB have better or similar performance than the Min,

and all ensemble models are better than the Tersoff. (b). Predicted elastic constants under Tersoff, RF, and AB versus DFT reference, both RF and AB models have lower MAEs than before. The points in the green circles have large errors, and all these points come from the complicated structures which are out of training sets in both ensemble models.

Table 1. MAEs and MADs of the partial elastic constants relative to DFT reference under AB and Tersoff. The smallest MAEs and MADs for each elastic constant are bolded for ease of reference.

| Methods | MAE of $C_{11}$ (GPa) | MAE of $C_{22}$ (GPa) | MAE of $C_{33}$ (GPa) | MAE of $C_{12}$ (GPa) | MAE of $C_{13}$ (GPa) | MAE of $C_{23}$ (GPa) | MAE of $C_{44}$ (GPa) | MAE of $C_{55}$ (GPa) | MAE of $C_{66}$ (GPa) |
|---|---|---|---|---|---|---|---|---|---|
| AdaBoost | **110** | **135** | **134** | **34** | **42** | **47** | **39** | **50** | **58** |
| Tersoff (with mp-1095534) | 1395 | 1732 | 156 | 1860 | 224 | 492 | 412 | 123 | 183 |
| Tersoff (without mp-1095534) | 266 | 262 | 157 | 138 | 68 | 65 | 100 | 107 | 152 |
| Methods | MAD of $C_{11}$ (GPa) | MAD of $C_{22}$ (GPa) | MAD of $C_{33}$ (GPa) | MAD of $C_{12}$ (GPa) | MAD of $C_{13}$ (GPa) | MAD of $C_{23}$ (GPa) | MAD of $C_{44}$ (GPa) | MAD of $C_{55}$ (GPa) | MAD of $C_{66}$ (GPa) |
| AdaBoost | **38** | **28** | **59** | **13** | **15** | **17** | **3** | **8** | **16** |
| Tersoff (with mp-1095534) | 291 | 302 | 173 | 75 | 102 | 38 | 116 | 246 | 206 |
| Tersoff (without mp-1095534) | 217 | 241 | 152 | 72 | 54 | 36 | 59 | 238 | 194 |

Besides, we combine formation energy and elastic constants data together to train and test the same four ensemble methods. The MAEs for both properties are shown in Fig. 4. All models perform weaker than before, with some MAEs even three times larger than when only one property is predicted. Even so, the MAEs of formation energy in RF, AB and GB models are lower than that calculated by over half of classical potentials, and the MAEs of formation energy in XGB are similar to AIREBO-M. Additionally, the MAEs of elastic constants in all models are lower than that calculated by all classical potentials, including Tersoff, though the MAE of Tersoff is lower if it removes the residuals of mp-1095534. The main reason for the increase in errors is the large size of the feature and the complexity of the correlation between features and targets will make regression trees hard to learn the relationship between features and targets correctly, the limited samples also prevent the models from learning the features of complex structures well. In our dataset, most structures are either graphitic or diamond-like. Graphite-like structures typically have anisotropic $C_{11}$, $C_{22}$, and $C_{33}$ elastic constants, with two of them usually close to 900 GPa, and their formation energy is lower than that of diamond-like structures, which is around 0.05 eV. On the other hand, the $C_{11}$, $C_{22}$, and $C_{33}$ are isotropic in diamond-like structures and all of them are around 1100 GPa, which are higher than those of graphitic structures, and have a higher formation energy of around 0.15 eV. Given these connections between formation energy and elastic constants, we can only use one type of property as a feature to reduce the dimension of the feature and predict both properties. For instance, when only applying the dataset of formation energy as features to train and predict both formation energy and elastic constants, we find that the MAEs are similar to those shown in Fig. 4.

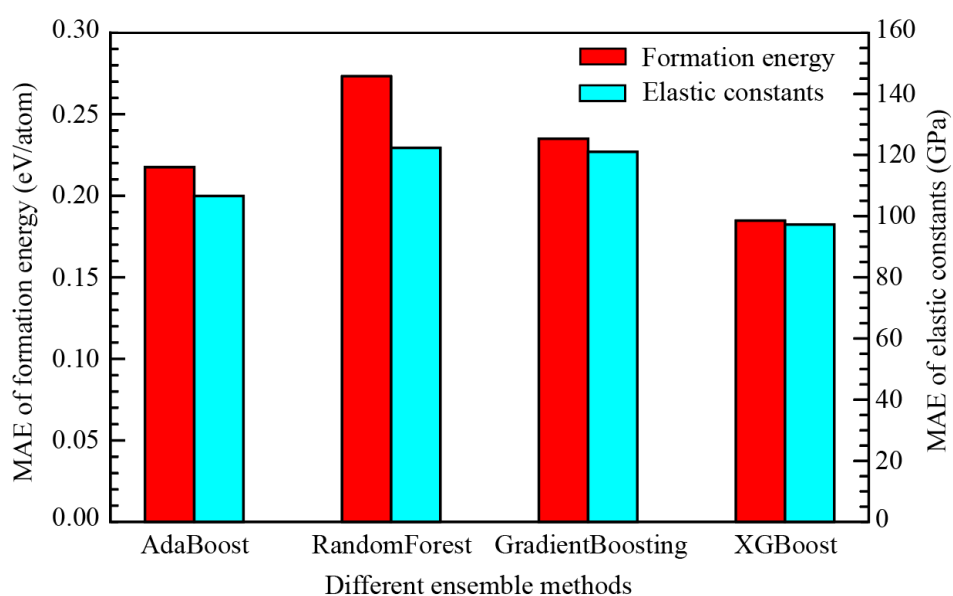

Fig. 4. MAEs of the formation energy and total elastic constants relative to corresponding DFT reference under different ensemble methods. Due to the complexity of the relationship between features and targets, all models perform weaker than when only one property is predicted. Even so, the MAEs of formation energy in RF, AB and GB models are lower than half of classical potentials. And the MAEs of elastic constants in all models are lower than that calculated by all classical potentials.

**Formation energy prediction of new carbon structures**

To further investigate the performance, the trained formation energy RF model is employed to predict the formation energy of new carbon structures. We extract all structures of both silicon carbide and silicon from the MP database and compared them with carbon structures by using the similarity method [59]. This method evaluates the dissimilarity of any two structures by calculating the statistical difference in local coordination environment of all sites in both structures. Out of 76 silicon carbide and silicon structures, 19 are extracted as new structures based on the similarity method. We replace the silicon element with the carbon element in these 19 structures to get new carbon structures and calculate their formation energy with the 9 classical potentials as features. These features are then input into the trained RF model to predict formation energy. Fig. 5a illustrates the formation energy of each new structure calculated by RF, DFT, and 4 classical potentials. For the first five structures with relatively low formation energy, RF model predictions are similar to those of DFT. But for other structures, the predictions are less accurate due to the limited training dataset (only 4 carbon structures in the training dataset have formation energy larger than 2 eV/atom).

These problems also exist in the classical potentials, especially for the structures with high formation energy that make the classical potentials hard to predict. Even so, the overall performance of RF is outperformed other classical potentials, as shown in Fig. 5b, the RF has the smallest MAE of all 19 structures. For the MAE of the first five structures, except larger than ReaxFF, RF has a similar MAE with ABOP, AIREBO, and AIREBO-M, and is smaller than the rest of the classical potentials. Interestingly, the LJ potential has relatively better performance than most potentials. Perhaps because the distance between carbon atoms in these structures is larger than that of the general carbon structure, causes the structure is unstable and the covalent bond is weakened, so that the non-bond potential such as LJ potential can better describe its energy.

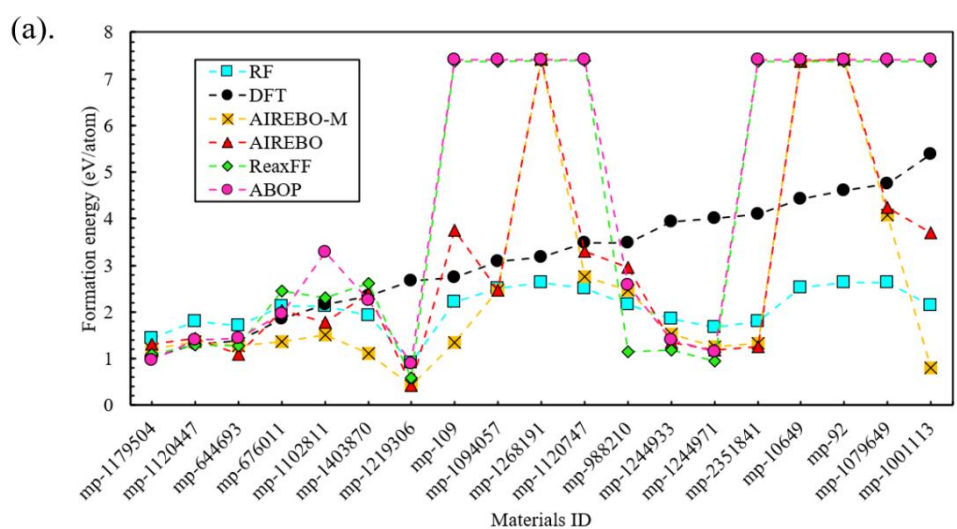

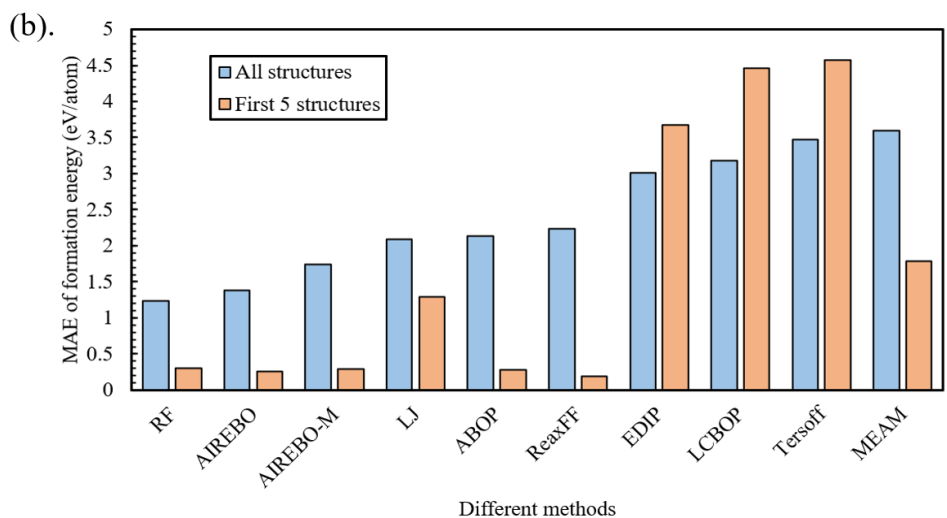

Fig. 5. Performance of different methods for prediction of formation energy of some carbon structures. (a). The formation energy of different carbon structures calculated by trained RF model and DFT. The x axis represents the ID of silicon carbide and silicon structures in MP database corresponding to the carbon structures. Due to the limited training dataset with formation energy greater than 2. The RF has weak performance for structures with high formation energy. (b). MAEs of formation energy of different methods. The RF has the smallest MAE overall, for the first five structures, RF has a similar MAE with ABOP, AIREBO, and AIREBO-M. The LJ potential has better performance than some classical potentials may be due to the structures being unstable and the covalent bond being weakened so that the non-bond potential such as LJ potential can better describe its energy.

**Interpretability**

To reveal the correlation or any useful information behind these features, principal component analysis (PCA) is used to decompose the high-dimensional dataset into a set of orthogonal components and project the dataset onto these components to indicate the maximum variance. Fig. 6a shows the projection of the 9-dimensional formation energy on a 2D plane. Four groups of carbon structures are labeled with different colors, and similar structures are clustered together compared to those labeled "others" with complex structures and mixed hybridization. In Fig. 6a, the graphite-like structures are grouped on the left of the plot, the diamond-like structures followed by fullerene-like structures are clustered on the right of graphite-like structures relative to the first principal component. And for the others, they are located scattered and on the right in general. This distribution is consistent with the formation energy of these structures that the graphite-like structures have the smallest formation energy followed by diamond-like and fullerene-like structures, and the complex structures have relatively higher formation energy. If combined with the second principal component, similar structures are close to each other, all of these indicate that the feature space contains the corresponding physical meaning that is consistent with that of the target property. These attributes are predictable since these features are the physical properties calculated by different containing intrinsic physical meaning classical interatomic potentials, so these features generally have similar values and distribution with the target properties.

Besides, the criteria for node splitting in regression trees is mainly based on the loss function like mean squared error (MSE), which describes the distribution of the targets under different features, and the regression trees will identify the feature with the minimal MSE as the threshold for the split point. Since there is a correlation between the features and targets in this study, it is ideal that the feature and corresponding targets have a linear relationship. So, the accuracy of the features varies for different structures, leading to different levels of linearity. Regression trees can evaluate the linear relationship of each feature at each split point and capture the most important feature that has minimal MSE. If a feature has relatively weak performance, it will have a nonlinear relationship between its values and targets. Thus, the targets corresponding to any two adjacent sorted values in this feature will become far apart, which causes the MSE to be larger compared to the more accurate feature. Fig. 6b shows the average feature importance of the regression tree fitted by formation energy, where 20 times permutation importance is employed for feature evaluation [46]. The permutation feature importance calculates the difference of error before and after permutation of the values of the features. In Fig. 6b, ReaxFF has the most impact on the accuracy of the model, followed by AIREBO-M. It should be noticed that LCBOP has a smaller MAE than ReaxFF and AIREBO-M in Fig. 2, this is because the node splitting depends on the linear relationship between features and targets instead of the difference between feature and target. Therefore, the Pearson correlation coefficient (PCC) is also used to assess the feature importance. PCC can measure linear correlation between two sets of data. The equation of PCC is as follows,

$$r = \frac{\sum(x_i - \bar{x})(y_i - \bar{y})}{\sqrt{\sum(x_i - \bar{x})^2 \sum(y_i - \bar{y})^2}} \tag{4}$$

Where $x_i$ and $y_i$ are values of the *x* and *y* variables, respectively, and $\bar{x}$ and $\bar{y}$ are their means. Fig. 6c is PCC of features and reference. From the last column of PCC, we can see ReaxFF has the largest positive linear correlation with the reference, and the regression tree will capture this linear correlation and use ReaxFF to split nodes, which indicates ReaxFF is the most important feature. The high positive linear correlation also explains why ReaxFF is more important than LCBOP although LCBOP has a smaller MAE in Fig. 2. However, the LCBOP's correlation with reference is higher than that of AIREBO-M, yet its importance is lower than that of AIREBO-M in Figure 6b. This suggests that other factors may also play a role in determining feature importance.

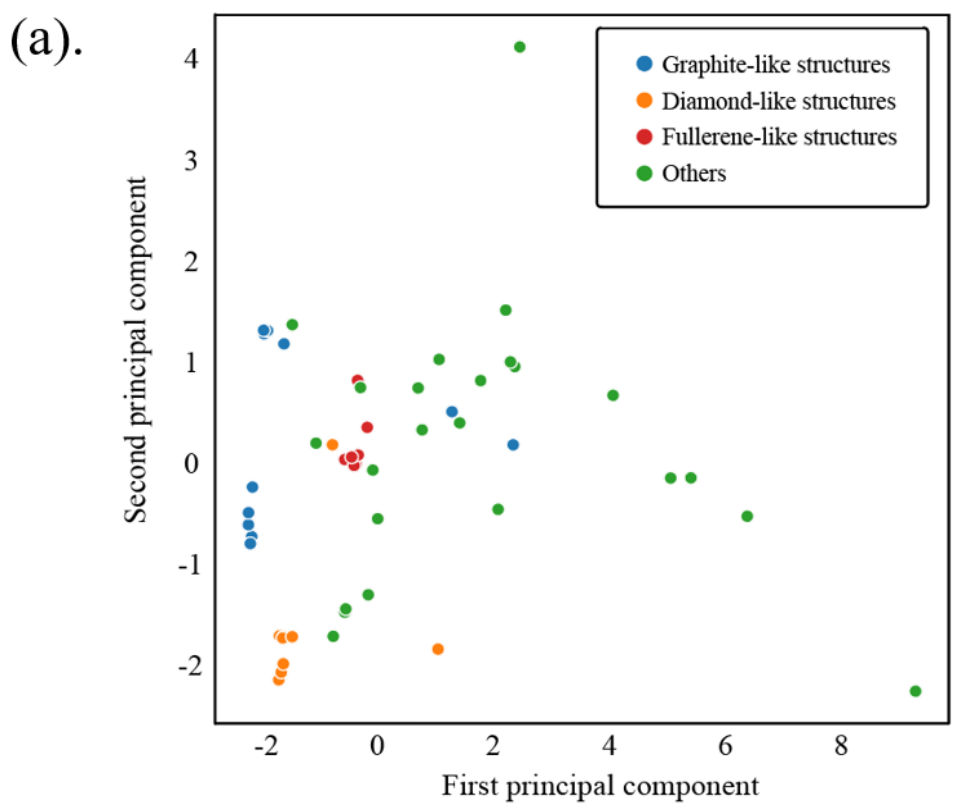

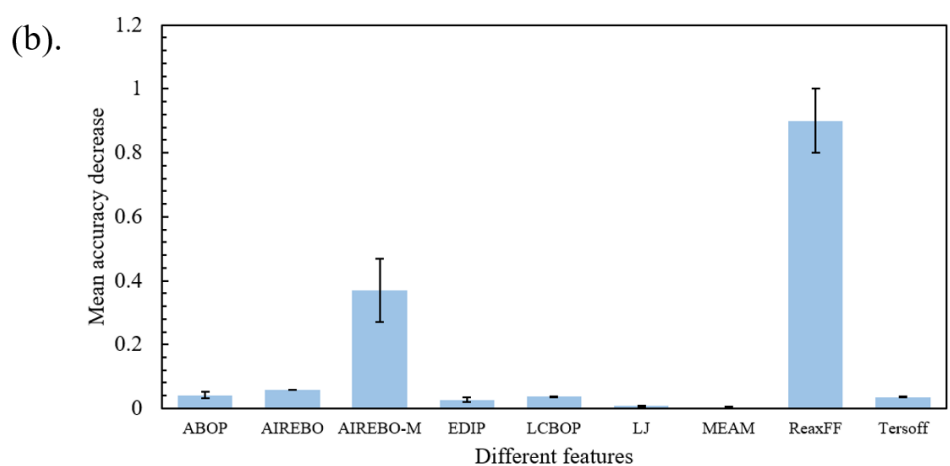

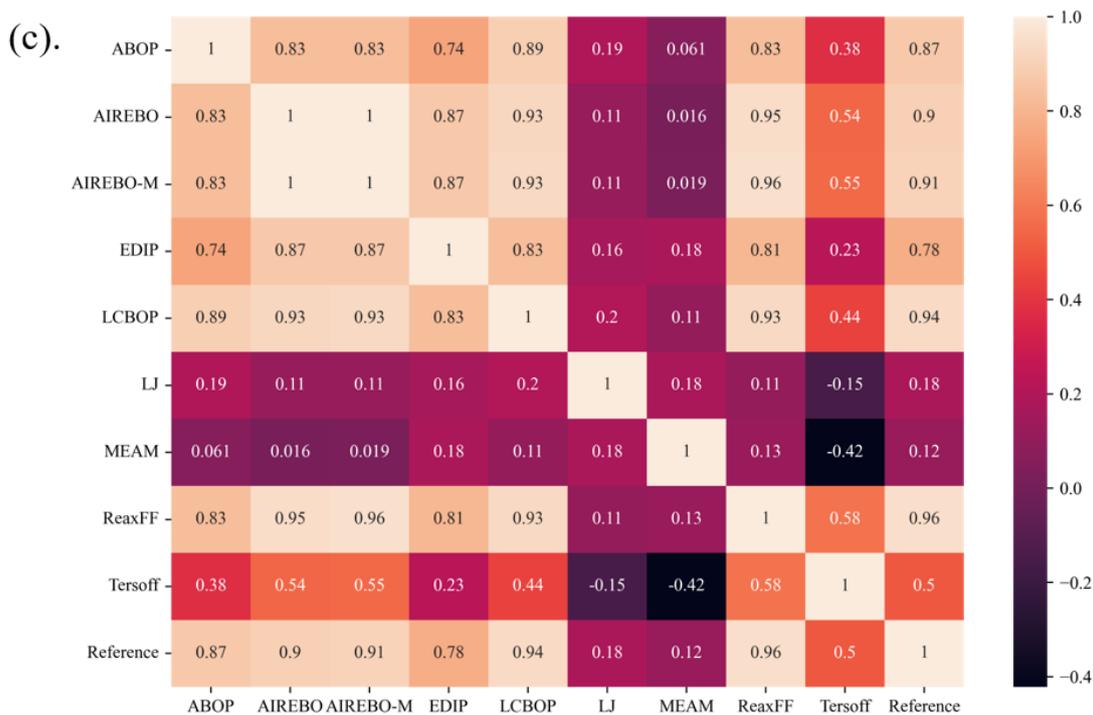

Fig. 6. Interpretability of feature importance with different methods. (a). Visualization of the features. The original 9D vectors are reduced to 2D with PCA. Similar structures are clustered together, and the others are scattered due to their different structure from each other. The distribution of the features for all structures on the first principal component is also similar to the distribution of the formation energy for the same structures. (b). Feature importance is calculated by permutating features. A higher value indicates a more important feature, ReaxFF and AIREBO-M are the two most important features for the accuracy of regression trees. (c). Pearson correlation coefficients of features and reference. Same as feature importance, ReaxFF has the largest PCC with the reference. The LCBOP, however, has a larger PCC than AIREBO-M, which means other factors may also play a role in determining feature importance.

Except for the factors mentioned before, the loss function indicates that local minimal error also impacts the choice of features for splitting. The regression tree algorithm is inherently greedy, aiming to find the feature with the local minimal error as a splitting feature if the sorted target values of the samples are close to each other. To quantify the level of local minimal error of each feature, we propose a way to describe the frequency of its occurrence. Fig. 7a illustrates the process of computing the local minimal level of each feature. In the beginning, initialize the local minimal level of each feature to 0, and split each feature's vector and target's vector into small parts, each part of the target consists of adjacent sorted target values and each part of the feature contains the MD-calculated values of structures corresponding to the target vector. For each part of the feature, we find its minimum and maximum values and compare them to each part of the target. If both minimum and maximum values exist in the target's part, which indicates a local minimal error and adds 1 to this feature's local minimal level. Fig. 7b shows the local minimal level of each feature, where ReaxFF has the largest local minimal level, followed by AIREBO-M. Thus, the local minimal level may explain a certain degree of the feature importance of ReaxFF and AIREBO-M. In summary, based on the linear correlation between features and targets and the local greedy characteristic of algorithms, ensemble learning can capture relatively accurate features calculated by the 9 classical potentials for splitting the node. To improve the performance of the ensemble model, more accurate features can be used.

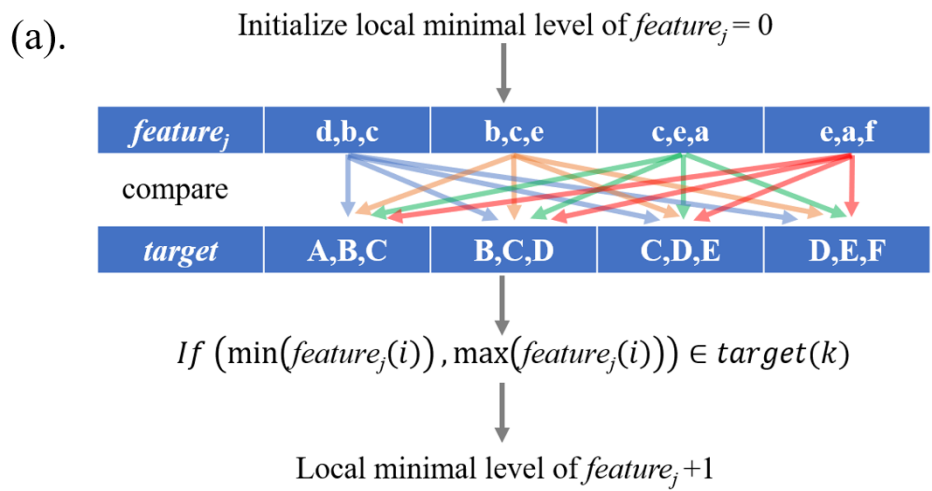

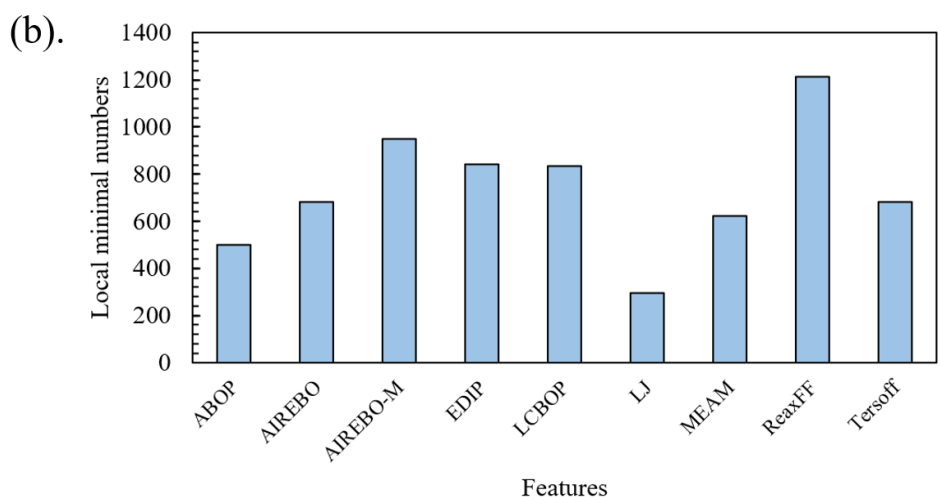

Fig. 7. Frequency of local optimization occurred in each feature. (a). Flowchart of calculating local minimal level of $feature_j$. In the beginning, initialize the local minimal level of each feature to 0 and split each feature's vector and target's vector into small parts, each part of the target consists of adjacent sorted target values (such as A-C represent three similar formation energy) in the figure, and each part of the feature contains the MD-calculated values of structures corresponding to the target vector (such as a-c). Then compare each feature's part to each target's part to check if both minimum and maximum values of each feature part also exist in target's part, if so, which means it has a local minimal error and add 1 to this feature's local minimal level. (b). Local

minimum level of different features. ReaxFF has the largest local minimal level, followed by AIREBO-M, which consists of the feature importance.

**DISCUSSION**

There are some limitations and opportunities for improvement. First, the limited size of the training dataset may restrict the performance of the models. This constraint is acknowledged in Fig. 3b, and it could be mitigated by including more training samples. Second, the performance of the regression trees is related to the accuracy of the features and a linear correlation between features and targets. So, more accurate classical interatomic potentials may be used as features to improve the performance. Third, more features make ensemble methods more complex and harder to interpret the feature importance, and complex correlations between features and targets also may lead to regression trees being unstable which affects the performance of ensemble learning, like the overall performance of the two properties prediction model (Fig. 4) is worse than single property prediction model (Fig. 2 and Fig. 3a), so to get better performance and interpretability, single property prediction model is preferred. Except for the limitations above, the input composed of physical properties calculated by different classical interatomic potentials is not convenient to get since it needs to calculate each new structure's physical properties by these potentials. However, inspired by the imputation of missing input values, this dilemma can be relieved by utilizing imputation methods to infer the missing values from the known part of the data. Here, we use k-Nearest Neighbors (KNN) approach [60] to impute the missing features in the input. KNN based on Euclidean distance metric to learn the correlation between features and find the nearest neighbors of the missing values among the samples that have values for the features, and the missing values are imputed using average weighted values from the nearest neighbors. Fig. 8 shows under the 10-fold cross-validation of the formation energy dataset, the performance of the RF model combined with 2-nearest neighbors' imputation when only one or two features use MD calculation. It can be clearly seen that when the more accurate features are calculated and other features are imputed, the accuracy of the model is higher. This is because the more accurate features are more important in the model, so calculating these features instead of imputing them will reduce the deviations of these features, thereby improving the prediction stability of the model. It can also be found from Fig. 8 that if more feature values are calculated as input, the accuracy of the model will be higher, such as when ReaxFF and AIREBO-M are used as input, the MAE is smaller than others, and the accuracy of the model is similar to that of the GB full-input model (Fig. 2). So, it is feasible to reduce the workload of obtaining the input part through the imputation of missing data though it will increase the error to a certain extent.

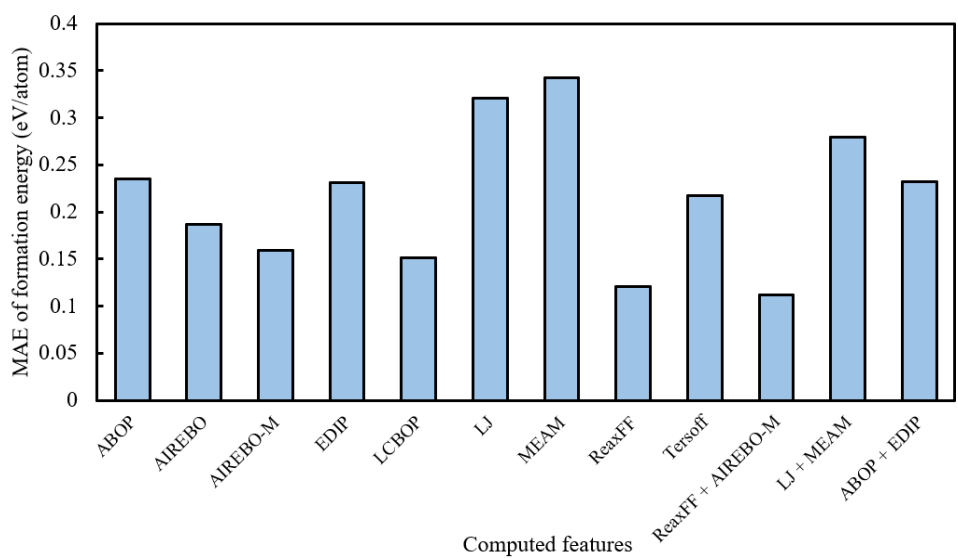

Fig. 8. The MAEs of formation energy of RF under 2-nearest neighbors' imputation. The x-axis represents different conditions, where the computed features are obtained by using MD simulations, and all other features in each condition come from 2-nearest neighbors' imputation. The RF model has better performance when more features or more accurate features are calculated as input instead of imputation.

In summary, we explore the possibility of prediction for the physical properties of a small size of carbon allotropes based on ensemble learning. The formation energy and elastic constants of carbon structures, as examples, can be predicted by using this kind of method. In general, the ensemble methods have better performance than the classical interatomic potentials we used in this work. Although at some points, the prediction is not accurate due to a lack of training data, the high dimensionality of features, and the local greedy characteristic of the algorithm, making the model difficult to learn the relationship between features and targets correctly. The principal component analysis shows the input which consists of the values calculated by different classical interatomic potentials, has a similar distribution with the corresponding target physical property. What's more, the Pearson correlation coefficient illustrates the linear correlation between input and output, and the regression trees can capture the relatively accurate feature as criteria for splitting the point in regression trees by evaluating the feature importance.

## Methodology

### Regression trees of ensemble learning

Regression trees, a type of decision tree, are used to predict outputs consisting of numerical values instead of categorical targets. They are also the base estimators in ensemble learning (the tree structures in Fig. 2). Fig. 9 illustrates a regression tree that has seven nodes in total. The tree starts from the top node, and each node contains sorted samples and will be split into two subsets based on the criteria (thresholds) for the features until the terminal condition is reached. The blue nodes are parent nodes, they have two subsets called children. The green nodes are end nodes, representing numerical outputs that are decided by the targets. In scikit-learn, the optimized version of Classification and Regression Trees (CART) [50] is used. This algorithm determines how to divide the sorted samples by trying different thresholds and calculating the mean squared error (MSE) at each step. In this study, the feature vectors $x_i \in R^n$ and target vector $y \in R^k$ are properties calculated by classical interatomic potentials and corresponding DFT reference, respectively. Where subscript $i$ represents indexes of different materials, superscript $n$ shows the number of input variables (the number of classical interatomic potentials), and superscript $k$ is the total number of materials. We denote $Q_m$ as the dataset at node m with $N_m$ samples, $Q_m^{left}$ and $Q_m^{right}$ as the children of $Q_m$, with $N_m^{left}$ and $N_m^{right}$ as the number of samples of these children. These children will split $Q_m$ into two parts using a threshold. The quality of the split of node m is calculated by minimizing the weighted average of impurity.

$$G(Q_m) = \frac{N_m^{left}}{N_m} H(Q_m^{left}) + \frac{N_m^{right}}{N_m} H(Q_m^{right}) \tag{5}$$

Where $H$ is loss function (such as mean squared error), for example, at node $m$, the mean squared error of its left child $Q_m^{left}$ is given by:

$$\bar{y}_m = \frac{1}{N_m^{left}} \sum_{y \in Q_m^{left}} y \tag{6}$$

$$H(Q_m^{left}) = \frac{1}{N_m^{left}} \sum_{y \in Q_m^{left}} (y - \bar{y}_m)^2 \tag{7}$$

Here, $\bar{y}_m$ is average value of target at node of $Q_m^{left}$. By recursing the $Q_m^{left}$ and $Q_m^{right}$, the weighted average of impurity changes as well, and selects the threshold that minimizes the impurity $G$ as the node $m$. Repeat and do the same steps for each node until the terminal condition is reached, and finally, a trained regression tree is obtained.

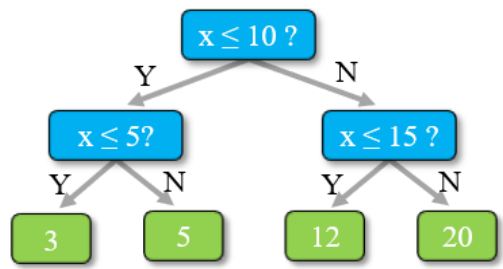

Fig. 9. The schematic of a regression tree illustrates the blue nodes (parent nodes) are split into two subsets (children) based on the threshold for the features until the terminal condition is reached. When using the tree to predict, the trained regression tree will depend on the inputs and the thresholds to select the children, finally, one green node

(output) will be decided.

**Bagging and boosting methods**

Bagging and boosting methods shown in Fig. 10 are used to achieve better performance than a single regression tree in this work. In bagging methods, several regression trees are trained independently by their own subsets in which the data can be chosen more than once. The final prediction is obtained by averaging the predictions [46] of all individual regression trees. On the contrary, the regression trees in the boosting method are generated sequentially and each regression tree has limited depth and is related to the previous one. Instead of averaging the outputs of all regression trees, the final prediction is calculated by calculating weighted median [47] of the predictions of all regression trees or summing predictions of all regression trees up [48]

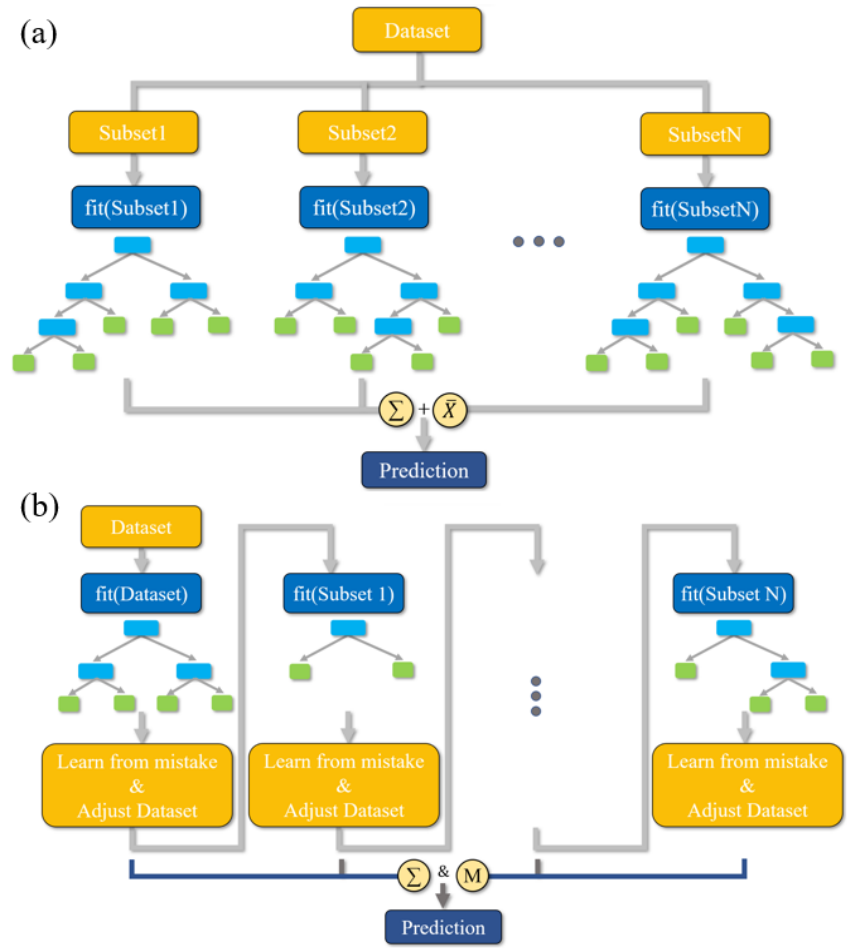

Fig. 10. Configurations of two regression-tree based ensemble learning models: bagging (a) and boosting (b). (a). In bagging methods, several regression trees are trained independently by their own subsets in which the data can be chosen more than once, and the final prediction is obtained by averaging the summation of all regression trees' predictions. (b). The boosting method is generated sequentially, and each regression tree is related to the previous one. The final prediction is weighted median of all predictions of regression trees or summation of all predictions of regression trees.

**DATA AVAILABILITY**

All data needed to evaluate the conclusions of this study are present in the paper, and all relevant data are available from Dr. Houlong Zhuang and Dr. Qiong Nian upon reasonable request.

**CODE AVAILABILITY**

All codes needed to evaluate the conclusions in the paper are available from Dr. Houlong Zhuang and Dr. Qiong Nian upon reasonable request.

**ACKNOWLEDGEMENTS**

Funding for this research was provided by National Science Foundation (NSF) under award numbers CMMI-1826439 and CMMI-1825739. This support is greatly acknowledged. The authors also thank the Agave Computer Cluster of ASU for providing the computational resources.

**COMPETING INTERESTS**

The authors declare no competing interests.